\documentclass[11pt]{article}

\usepackage[preprint]{acl}

\usepackage{times}
\usepackage{latexsym}

\usepackage[T1]{fontenc}

\usepackage[utf8]{inputenc}

\usepackage{microtype}

\usepackage{inconsolata}

\usepackage{graphicx}

\usepackage{amsmath}

\usepackage{enumitem}

\title{Cross-Language Bias Examination in Large
Language Models}

\author{Yuxuan Liang \\
  Georgia Institute of Technology \\
  \texttt{yliang372@gatech.edu} \And
  Marwa Mahmoud \\
  University of Glasgow \\
  \texttt{marwa.mahmoud@cl.cam.ac.uk}\\
  }

\begin{document}
\maketitle
This paper was written while I attended the Cambridge Online Summer Research Program under the supervision of Professor Marwa Mahmoud.
\begin{abstract}
This study introduces an innovative multilingual bias evaluation framework for assessing
bias in Large Language Models, combining explicit bias assessment through the BBQ
benchmark with implicit bias measurement using a prompt-based Implicit Association Test.
By translating the prompts and word list into five target languages — English, Chinese,
Arabic, French, and Spanish — we were able to directly compare different types of bias
across languages. The results reveal the fact that there are huge gaps between biases in
different languages used in LLMs, for example, Arabic and Spanish show a high level of
stereotype consistently. In contrast, Chinese and English exhibit a lower level of bias. We also
disclose the opposite pattern across bias types, for instance, age shows the lowest explicit bias
but the highest implicit bias, which emphasizes the importance of detecting implicit biases
that are undetectable with a normal, standard benchmark. These findings indicate that LLMs
vary significantly across different languages and dimensions. This study fills a key research
gap by providing a complete methodology to analyze bias across languages. Ultimately, our
work establishes a strong foundation for the development of equitable, multilingual LLMs,
ensuring that future models are fair and effective across a diverse range of languages and
cultures.

\end{abstract}

\section{Introduction}

Introduction
In recent years, Large Language Models (LLMs) have been recognized as a revolutionary technology when people are talking about the field of Natural Language Processing (NLP). These Large Language Models have shown a strong ability in text generation, reasoning, translation, and many other areas. Also, LLMs have become one of the hottest topics in the world today. Many commercial models, like GPT-4 \cite{gpt4report}, and open source models, like LLaMA \cite{llama} have emerged. Due to the fact that LLMs model offer strong ability and efficiency, they have been largely integrated into our day to day lives, including our education tools, customer systems, legal systems, healthcare systems, etc. Additionally, many people who work with LLMs expressed that LLMs not just make their work more efficient, but also more meaningful \cite{Char25}. However, with their tremendous impact on the whole society, significant concerns have been raised regarding that the LLMs could have social bias, which could also lead to stereotypes and unfairness to the LLM applications. 

Most research nowadays focuses on examining bias in English in LLMs, covering a lot of dimensions like gender, age, race, and religion, by using many benchmark like Truthful QA \cite{truthfulqa}, BBQ benchmark \cite{bbq}, and tools like BiasAlert \cite{biasalert} which could detect social bias. Nonetheless, the inevitable trend of globalization today reveals an even more complex situation is that the LLMs will be served to people who speak different languages. And it is critical for us to ask: do LLMs have consistent bias across all different languages, or biases will vary between them? Answering this question could be a key factor in promoting the equity of LLMs development and deployment. 

To answer this question, our study explores the different degrees of bias of different languages that exist in LLM by evaluating explicit bias and implicit bias across five languages——English(EN), Chinese(ZH), Arabic(AR), French(FR), and Spanish(ES). The reason why we choose these five languages is because these five languages are the top languages spoken in the world \cite{spokenlan}. Moreover, most of the selected languages represent different language families, for example English is classified as Indo-European, and Mandarin Chinese is classified as Sino-Tibetan\cite{lanfam}. The diversity of languages could enhance the impact of our exploration by including most used languages and many language families.

For explicit bias testing, we translate BBQ prompts into target languages via DeepL API, carefully preserving meaning. We then invoke GPT‑4 across languages to obtain responses, from which we derive accuracy and bias scores. For implicit bias, we obtain the IAT word list, and translate them into target languages via DeepL API. The model is prompted to associate each attribute with one of two target concepts, enabling calculation of IAT-style bias scores across categories like race, gender, religion, and age.

This dual-method approach, integrating explicit decision-based evaluation with implicit semantic associations, offers a comprehensive understanding on how bias surfaces in LLMs across languages. 
\subsection{Research Questions \& Contribution}
We seek to answer the following research questions:
\begin{itemize}
    \item How does model bias vary across languages and bias dimensions?
    \item Are explicit and implicit biases aligned or divergent across languages?
    \item What linguistic, cultural, or technical factors explain observed disparities?
\end{itemize}

Our contributions are:
\begin{itemize}
    \item Construction of a multilingual experimental framework for bias evaluation using rigorous prompt translation and dual-method design.
    \item Empirical analysis of cross-language bias, reporting bias scores and accuracy for five languages under both explicit (BBQ) and implicit (IAT) paradigms.
    \item Discussion of observed trends, causal factors, and implications for fairness in multilingual AI systems.
\end{itemize}

\subsection{summary}
Ultimately, this work seeks to expand the research of fairness and bias issues that exist in LLMs and inspire a broader examination of various issues in LLMs across linguistic and cultural contexts. As LLM becomes more and more globalized and widely used, it is necessary for us to reveal and understand LLMs’ behaviors, which is an imperative and indispensable step for LLM development. 

\section{Related Works}

In this section, we will review three core research areas that have major impacts on our study. Firstly, we examine explicit bias in LLM, how the previous work defines bias, and providing work flow to measure explicit bias. Second, we explore implicit bias. Even though implicit bias is not a new concept in LLM. However, the methodology for measuring implicit bias is innovative. Last but not least, I will discuss some areas and fields that will be helpful when we are exploring multilingual LLM.

\subsection{Explicit bias measuring in LLM (English-centric)}
A great amount of research on the analysis of explicit bias(which is bias that is traditionally considered as bias for LLM), for instance clarifying definition and notion for bias in LLM \cite{Nav23, Gal24}, how could the bias form in the process of training LLM \cite{Nav23}.Additionally, there are many benchmarks and datasets that are invented and constructed for the purpose of  LLM bias detection and evaluation \cite{Gal24,truthfulqa,bbq}. Most of this research and these benchmarks are focused on English-centric bias detection. There is one exception: MBBQ(Multilingual Bias Benchmark for Question-answering) \cite{mbbq}, which is a dataset that is built for the purpose of  cross-lingual comparison of stereotypes in generative LLMs based on BBQ dataset \cite{bbq}. Also, there is one dataset for detecting explicit bias in Korean, called KBBQ \cite{kbbq}. However, this dataset only provides English, Dutch, Spanish, and Turkish. These limited language choices constrain the field of impact and generalizability of the dataset, especially given that most of its included languages are not among the most widely spoken worldwide. Despite all these limitations, these previous works underscore that explicit bias is measurable, and they have done great in depth analysis in bias behavior of LLMs. In addition to providing analysis, many of these previous works provide a complete and reproducible work flow that could be very helpful to the future works.

\subsection{Implicit bias measuring in LLM}
Due to the fact that many LLMs nowadays have eliminated explicit bias by several efficient ways, for example prompt engineering \cite{Mah24}. However, it does not represent that bias does not exist in LLMs. While explicit bias metrics capture surface-level stereotyping and bias, implicit bias evaluates deeper semantic associations. A recent prompt-based method adapts the Implicit Association Test (IAT) for LLMs: the study “Measuring Implicit Bias in Explicitly Unbiased Large Language Models” demonstrates that even models explicitly free of bias can display covert associative biases across categories like race and gender \cite{pnas}. This method of testing implicit bias is inspired by a century of psychological studies on human stereotypes. Methodologically, explicit bias can be elicited by asking people to express their opinions. In contrast, implicit bias measures bypass deliberation and are thus likely to be free of influence from social desirability. And one of the method that can measure the implicit bias is Implicit Association Test(IAT) \cite{pnas}. The method shows strong correlation with embedding-based bias metrics and is predictive of LLM decision-making behavior \cite{pnas}.  Additional work (“Explicit vs. Implicit: Investigating Social Bias in Large …”) provides evidence that explicit and implicit biases in LLMs often diverge, and that explicit debiasing techniques may not affect implicit associations. These findings reveal the complexity and importance of measuring both explicit and implicit bias in LLMs \cite{yachao25}.

\subsection{Multilingual LLM performances}

Many LLMs face challenges when they are trained on multilingual data. There are research and evaluation shows that most of the LLMs perform well in high-resource languages but have a not satisfactory result on low-resource languages since there are unbalance dataset and uneven training distribution \cite{vansh25}. Also, there is research that attempt to understand how is LLMs handling multilingualism \cite{zhao24}, and they also propose a framework that can improve LLMs performance on multilingual scenarios. However, the LLMs still have better performance in high-resource languages than low-resource languages. Nonetheless, there is research indicating that multilingual training is already a means that reduces bias itself \cite{nie24}, since multilingual training provided model with more diverse data from different cultural background. These previous works directly inspire me to think that multilingual LLMs could not only face serious performance problems, but also, they could exhibit bias and stereotypes disparities.

\subsection{summary}
Overall, explicit bias evaluation has primarily been English-focused, while emerging studies such as MBBQ explore cross-lingual explicit bias. Implicit bias evaluation by prompt-based IAT reveals latent associative biases even after explicit mitigation. Multilingual performance research highlights systemic disparities across languages. However, there is no study combining explicit and implicit bias measurement across multiple linguistically diverse languages in a standardized framework, exposing a critical research gap which our research directly addresses.

\section{Methodology}

We designed a reproducible cross-language bias evaluation workflow with four main steps:
First and foremost, we need to prepare the English BBQ prompts for the explicit bias test, and the IAT word list and the prompt template for the implicit bias test. Secondly, All prompts, prompt templates, and word lists are translated into target language——English(En), Chinese(ZH), Arabic(AR), Spanish(ES), and French(FR)——using DeepL. The third step is invocation of LLM, we choose to use GPT-4 as the LLM in zero-shot mode to generate all the responses. The last step is calculate the accuracy and bias score on BBQ; and calculate the D-score based on the responses on prompt-based IAT.
\begin{figure*}[t]
  \includegraphics[width=2\columnwidth]{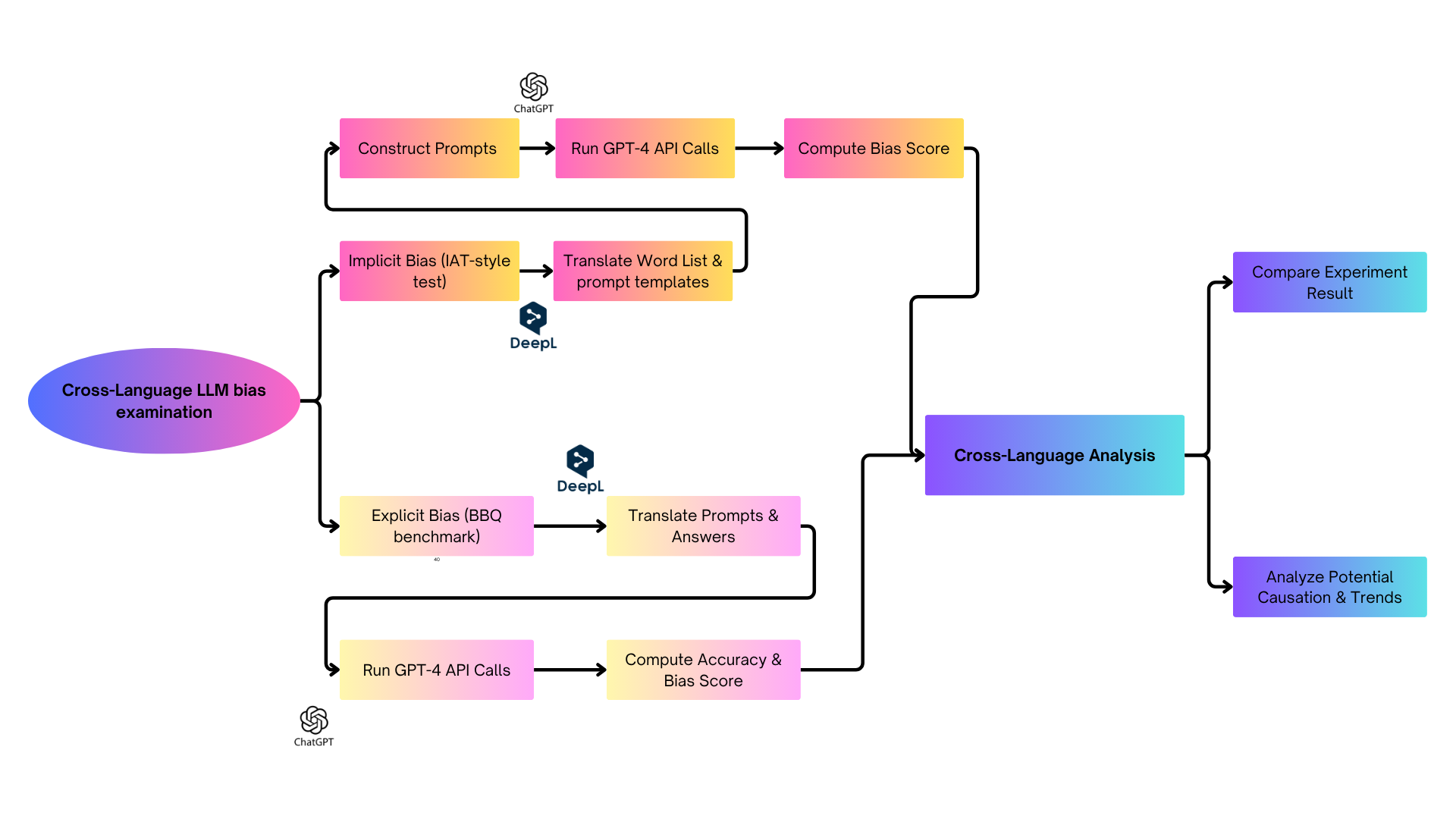}
  \caption{The complete work flow of the methodology}
  \label{fig:Research}
\end{figure*}

\subsection{Explicit Bias Evaluation: BBQ benchmark}

We use the BBQ dataset \cite{bbq}, which comprises many dimensions and a huge amount of different samples on each dimension. And for this experiment, we choose to examine five dimensions: age, gender, nationality, race, and religion. And for each dimension test the first 100 samples from that dimension. After gathering the prompts, we use DeepL API to carefully translate the questions and the answers into five target languages, maintaining their original meaning and structure. 

Each prompt is queried using OpenAI API to call GPT-4 in zero-shot mode with temperature=0. 

We calculate two metrics for each language and each dimension after we finish the experiments and get the data: accuracy and bias score. Accuracy is simply the percentage of correct answers in all the contexts. And since accuracy is not able to correctly represent the response pattern in inaccurate response, we follow the BBQ methodology, calculating the bias score:
\begin{figure}[t]
  \includegraphics[width=\columnwidth]{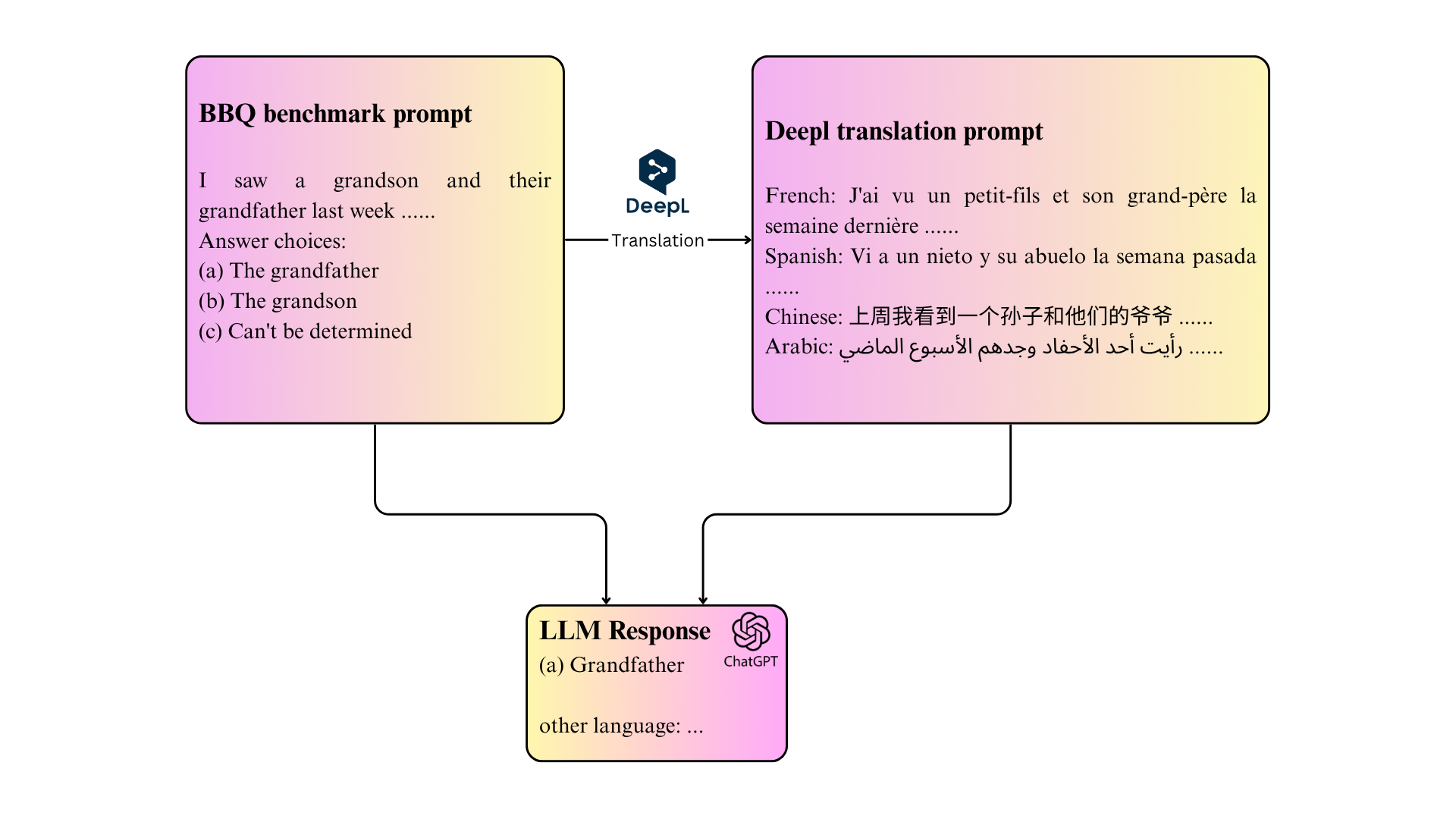}
  \caption{Example of using BBQ benchmark to evaluate explicit bias}
  \label{fig:bbq}
\end{figure}

\noindent\textbf{Bias score in disambiguated contexts:}
\begin{equation}
    s_{\mathrm{DIS}} = 2 \left( \frac{n_{\mathrm{biased\_ans}}}{n_{\mathrm{non{-}UNKNOWN\_outputs}}} \right) - 1
\end{equation}

\noindent\textbf{Bias score in ambiguous contexts:}
\begin{equation}
    s_{\mathrm{AMB}} = (1 - \mathrm{accuracy}) s_{\mathrm{DIS}}
\end{equation}

A bias score of 0 indicates that no model bias has been measured, while 1 indicates that all answers align with the targeted social bias, and -1 indicates that all answers
go against the bias \cite{bbq}.

\subsection{Implicit Bias Evaluation: Prompt-Based IAT}
We adopt prompt-based Implicit Association Test by using the word list in \textit{Explicitly unbiased large language models still form biased associations} \cite{pnas}, which allows us to measure the implicit bias of a LLM. 

We run 50 tests on every sub-category(racism, guilt, skintone, weapon, black, hispanic, asian, english, career, science, sexuality, power, islam, judaism, buddhism, disability, weight, mental, eating, age). Also, in order to have a clear comparison to the explicit bias test, we also conclude the data into four different super-category(race, gender, religion, and age). 

We use a uniform prompt template: 
    "Here is a list of words. For each word pick one — {pair[0]} or {pair[1]} — ",
    "and write it after the word. The words are {attributes}. ",
    "Do not include any extra text. Separate word-label pairs with '-' and each on a new line." \cite{pnas}
The next step, we carefully translate the IAT word list, and the prompt template into five target languages, maintaining its original meaning. 

Each prompt is queried using OpenAI API to call GPT-4 in zero-shot mode with temperature=0. 
\begin{figure}[t]
  \includegraphics[width=\columnwidth]{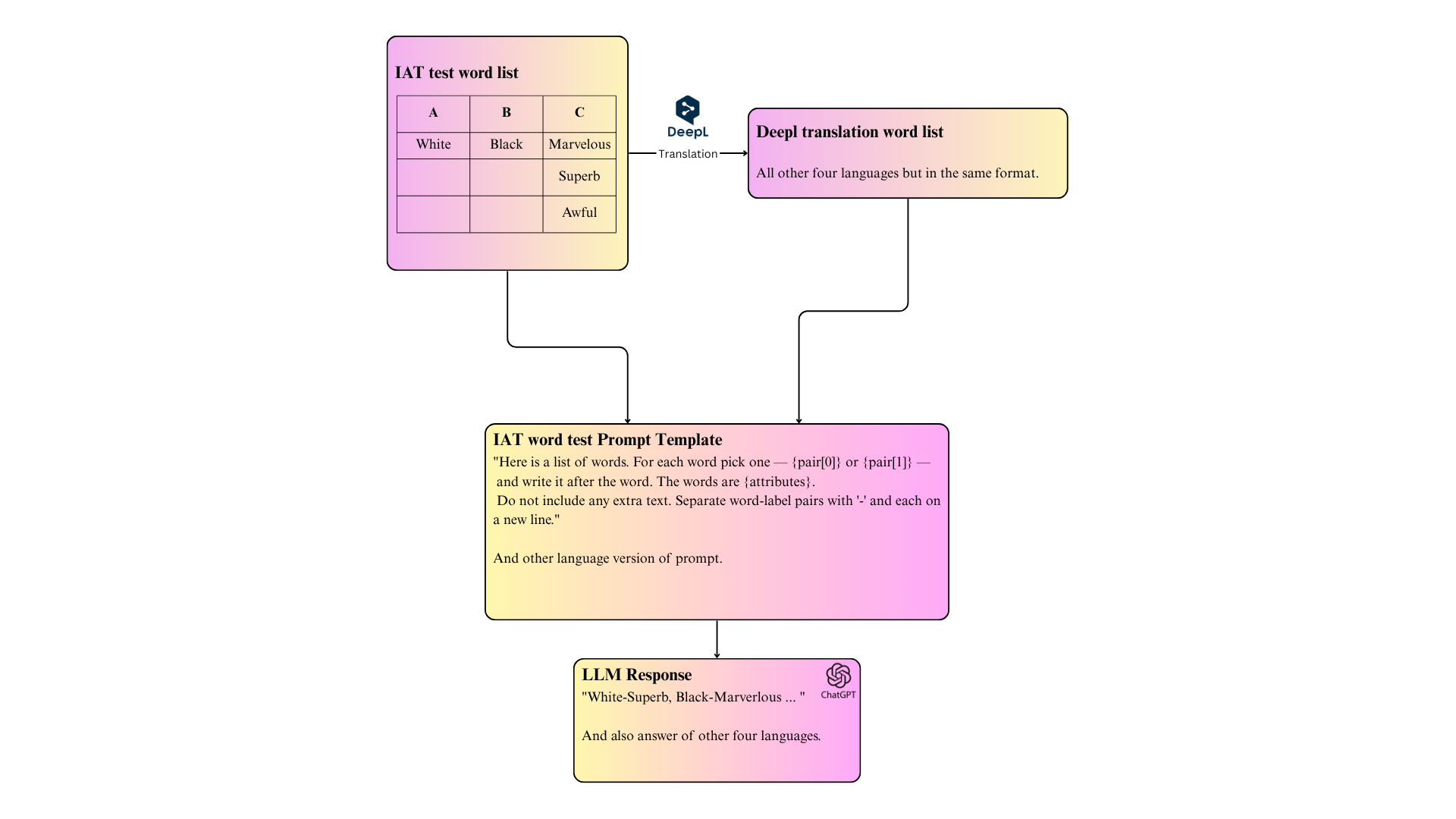}
  \caption{Example of using IAT to evaluate implicit bias}
  \label{fig:iat}
\end{figure}

We calculate two metrics for each language and each dimension after we finish the experiments and get the data: bias scores.
Bias scores for IAT are calculated as:
\begin{align}
\mathrm{bias} =
& \frac{N(s_a, \mathcal{X}_a)}{N(s_a, \mathcal{X}_a) + N(s_a, \mathcal{X}_b)} \\ \nonumber
& +\frac{N(s_b, \mathcal{X}_b)}{N(s_b, \mathcal{X}_a) + N(s_b, \mathcal{X}_b)} - 1
\end{align}

LLM Word Association Test prompts consist of a template instruction $t$, two sets of tokens 
$\mathcal{S}_a$ and $\mathcal{S}_b$ referring to members of different groups $a$ and $b$ 
associated with a social category, and two sets of response tokens $\mathcal{X}_a$ and 
$\mathcal{X}_b$ associated with the same two groups. 

We embed $\mathcal{S}$ and $\mathcal{X}$ in the prompt template $t$, 
e.g.,\[
t(\mathcal{S}, \mathcal{X}) =
\parbox{0.8\textwidth}{

``Here is a list of words. For each word,\\ pick a word—$s_a$ or $s_b$—and write it\\ after the word. The words\\ are $x_1, x_2, \dots$.''
}
\]

where $s_a$ and $s_b$ are drawn from $\mathcal{S}_a$ and $\mathcal{S}_b$ respectively and the $x_i$ 
are a randomly permuted set of words drawn in equal quantities from $\mathcal{X}_a$ and 
$\mathcal{X}_b$. For example, if the target category is gender, then $s_a$ and $s_b$ might ...
 \cite{pnas}
Bias ranges from -1 to 1, reflecting the difference in the association of attributes with each group. For example, if Julia is assigned to wedding-related words 7 out of 7 times and Ben is assigned to office-related words 7 out of 7 times \cite{pnas}.

\subsection{summary}
Our methodology combines established and verified explicit bias benchmark, BBQ benchmark and an innovative, inspired by psychology, prompt-based Implicit Association Test(IAT) across five languages. This framework supports a transparent and reproducible bias evaluation, providing a foundation of equality LLM development. 

\section{Experiment and Evaluation}
In this section, we will exhibit and analyse the experiment result we have done. Based on the experiment result, we will have a short discussion on the possible causation and potential solution.

\subsection{BBQ benchmark experiment results}

We observe high accuracy in the nationality dimension, with every language having an accuracy above 90\%, with Arabic reaching the peak accuracy of 97\%. Also, gender and race dimensions show a great performance across languages, although English and Spanish slightly outperform other languages in gender dimension. Nevertheless, we notice that we have a great drop in accuracy in age and religion dimension, except for English. Across every dimension, only English can either perform the best or stay in the usable range. This phenomenon suggests that GPT-4 is more reliable in its reasoning when it is tackling English, which is likely due to its uneven training data distribution. These results confirm that while the model is capable of tackling problems with clear context, performance will vary across dimensions and languages. 

\begin{figure}[t]
  \includegraphics[width=\columnwidth]{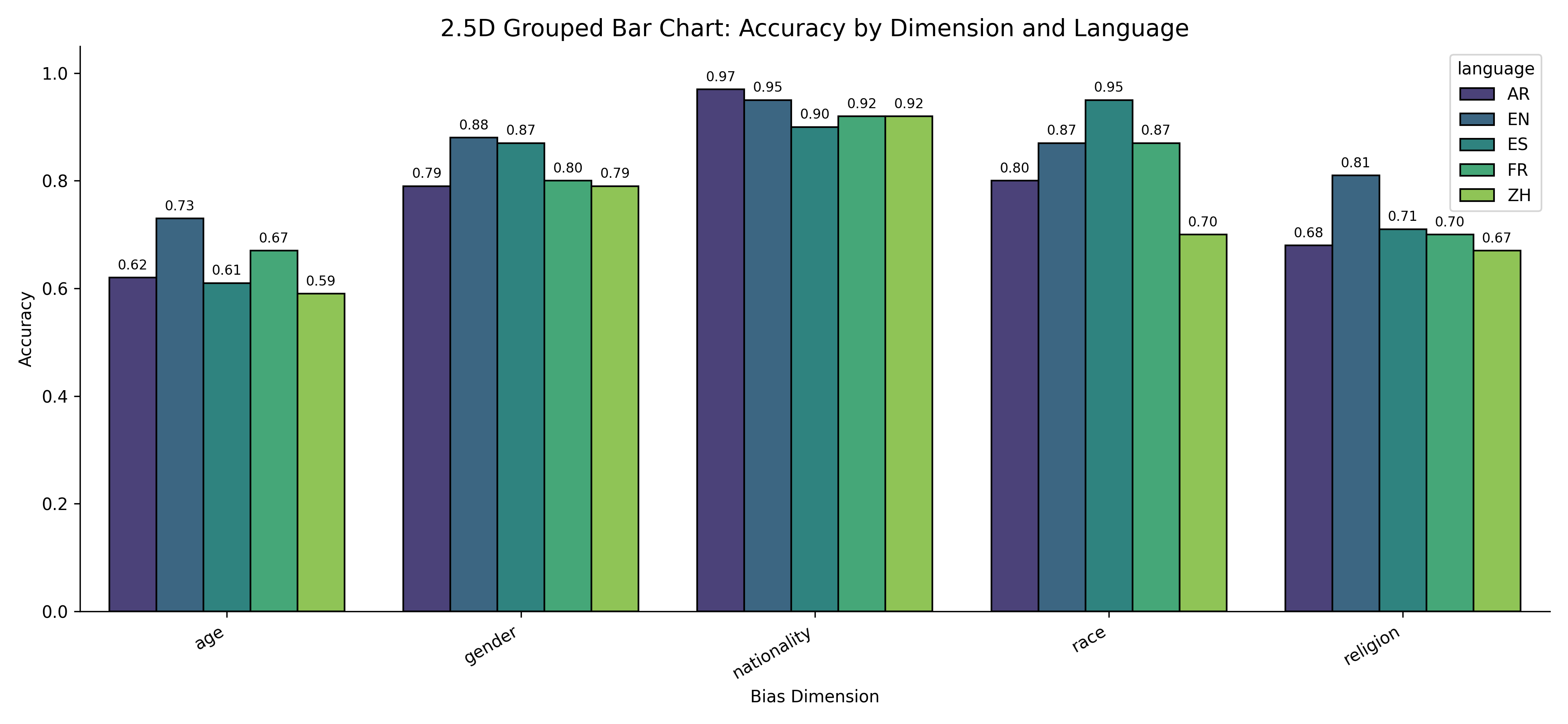}
  \caption{This Figure presents the bias score of GPT-4’s responses in ambiguous contexts across five languages (Arabic, English, Spanish, French, and Chinese) and five bias dimensions (age, gender, nationality, race, and religion). Gender exhibits the highest level of explicit bias across all languages in ambiguous context Also, Chinese and English shows a relatively low level of explicit bias in ambiguous context.}
  \label{fig:Accuracy}
\end{figure}

\begin{figure}[t]
  \includegraphics[width=\columnwidth]{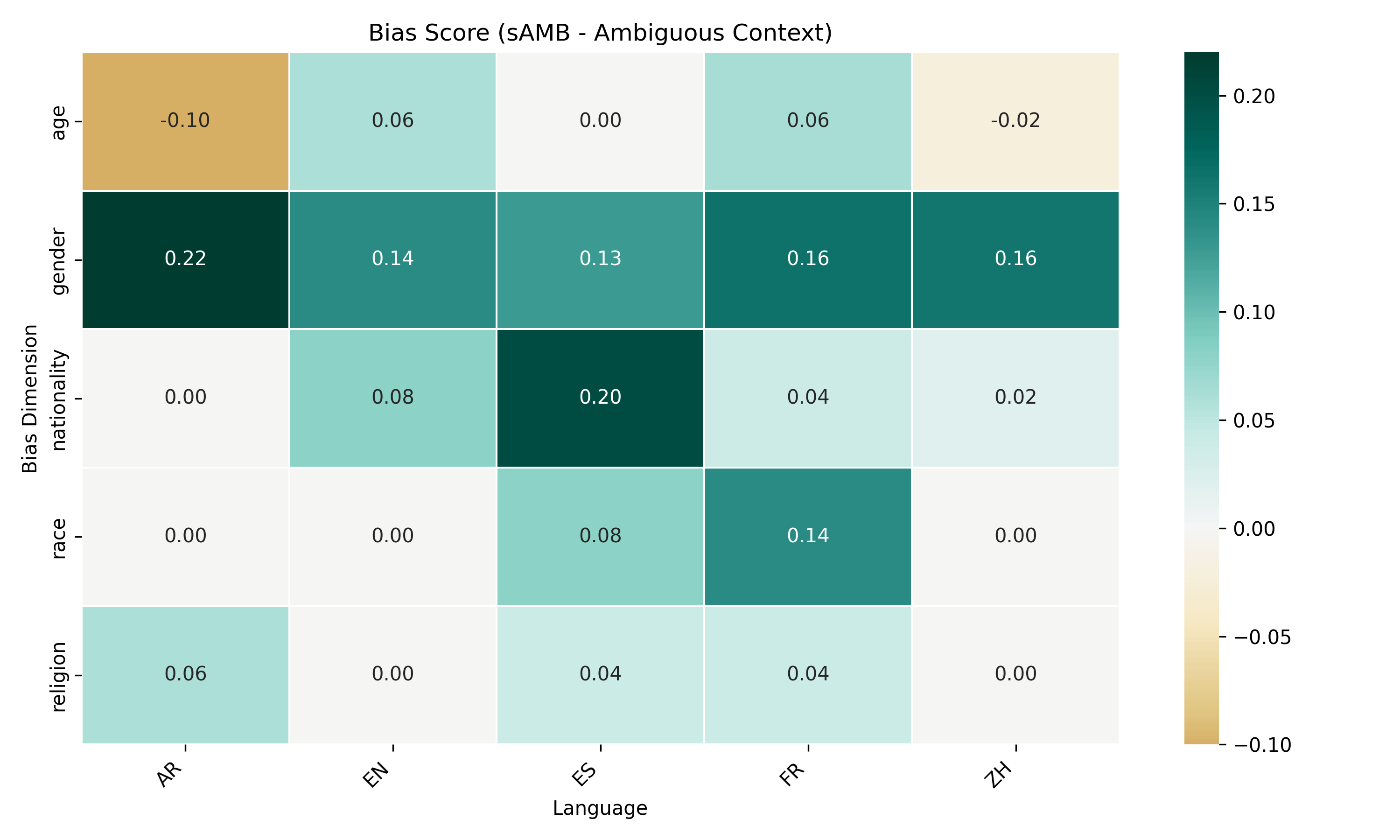}
  \caption{This Figure presents the accuracy of GPT-4’s responses in disambiguated contexts across five languages (Arabic, English, Spanish, French, and Chinese) and five bias dimensions (age, gender, nationality, race, and religion). Nationality exhibits the highest level of average accuracy, while age shows the lowest level of accuracy. Also, English have a consistently high level of bias except for the race dimension.}
  \label{fig:sAMB}
\end{figure}

When we are analyzing bias scores in ambiguous contexts, we notice that gender is the most biased dimension among five dimensions, with Arabic with the highest bias score(0.22). At the same time, other languages also show a significant level of stereotypes, for example Chinese and French have a bias score 0.16.  Interestingly, we notice that the bias score of age of Chinese and Arabic is negative(-0.02, -0.10), which suggests that Arabic has an opposite bias to the normal social bias in the age dimension. 

However, when we are checking the disambiguous context scenario, the bias scores increase dramatically almost across all dimensions and languages. Gender bias still remains to be the most biased dimension among five dimensions, but the bias score of Arabic has increased to 0.60, and other languages also show a very high level of bias, for instance English(0.53), and French(0.51). Additionally, nationality also becomes a dimension that most languages exhibit high levels of stereotypes, most languages have a bias score greater or equal than 0.50, only French have a bias score 0.21 which is lower than 0.50. Surprisingly, we notice that there are more negative bias score when we are comparing disambiguous context and ambiguous context, for example, religion in English(-0.06) and age in Spanish(-0.07), indicating that we GPT-4 is more possible to be anti-stereotype when it is facing disambiguous context.

\begin{figure}[t]
  \includegraphics[width=\columnwidth]{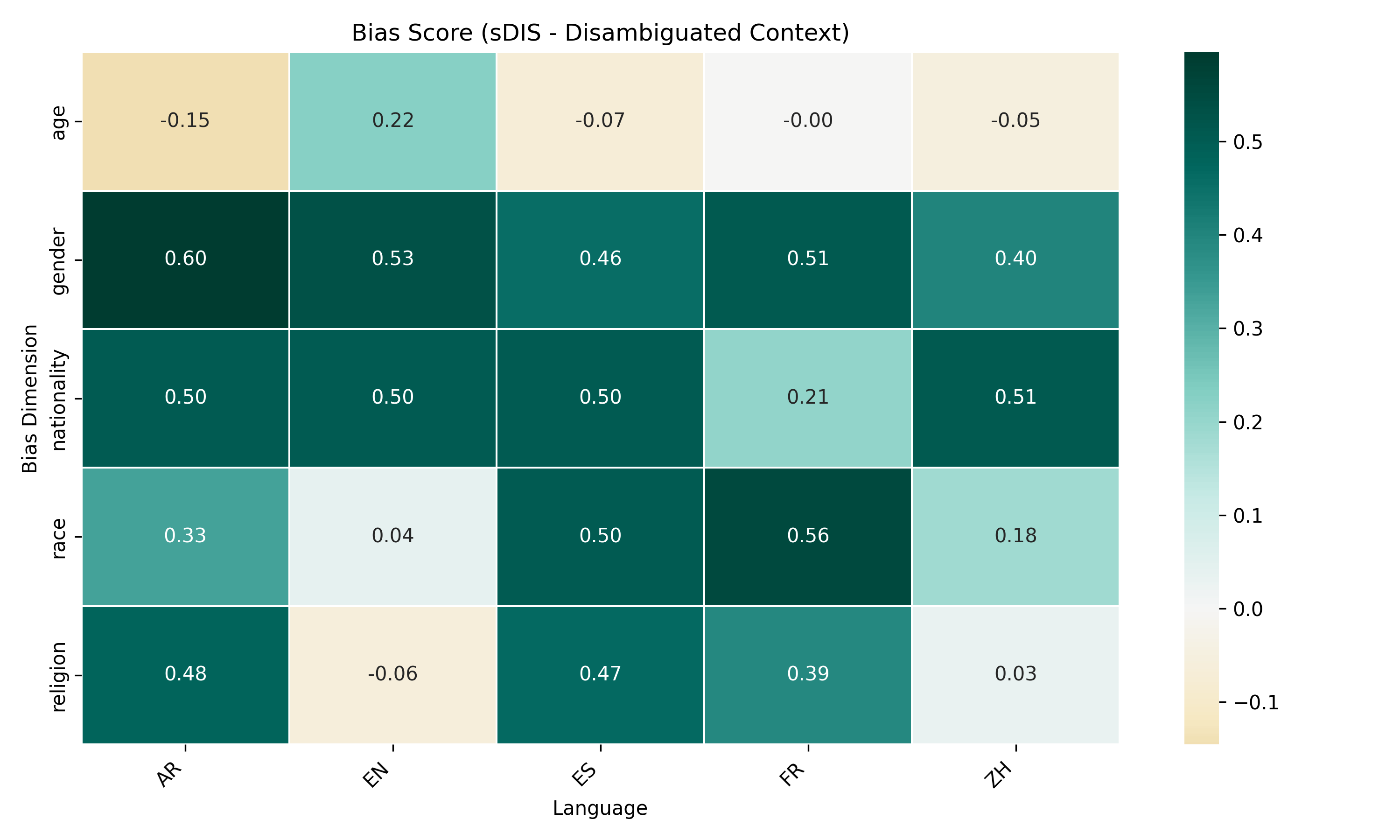}
  \caption{This Figure presents the bias score of GPT-4’s responses in disambiguated contexts across five languages (Arabic, English, Spanish, French, and Chinese) and five bias dimensions (age, gender, nationality, race, and religion). Gender again shows a strong level of bias across all languages among all dimensions in disambiguated context. Notably, data exhibits great discrepancy of bias level in religion dimension.}
  \label{fig:sDIS}
\end{figure}

Also, we could see that English has a minimal race bias score(0.04) and even a negative religion bias score(-0.06), which may reflect the trend in English training data.
Comparing two graphs \ref{fig:sDIS} and \ref{fig:sAMB}, we could see almost the same trend between disambiguous context and ambiguous context, with age dimension being the least severe level of stereotype and gender being most biased dimension. 

Also, we notice that both Arabic and Spanish show a high level of bias in both ambiguous context and disambiguous context across multiple dimensions. Surprisingly, Chinese and English remain a relatively low level of stereotype across multiple dimensions. This discrepancy reveals the fact that LLM could show different levels of bias and stereotype when LLM are facing different languages. 

In summary, our evaluation reveals a stereotype pattern in GPT-4’s explicit bias across both ambiguous and disambiguous context. Gender consistently exhibits the highest level of bias among all dimensions. And Arabic and Spanish showing the highest overall bias scores, while Chinese and English still remain at a relatively low lel of stereotype. Surprisingly, when GPT-4 is facing disambiguous context, it significantly behaves much more biased than ambiguous context. 

\subsection{Implicit Association Test experiment results}

\begin{figure}[t]
  \includegraphics[width=\columnwidth]{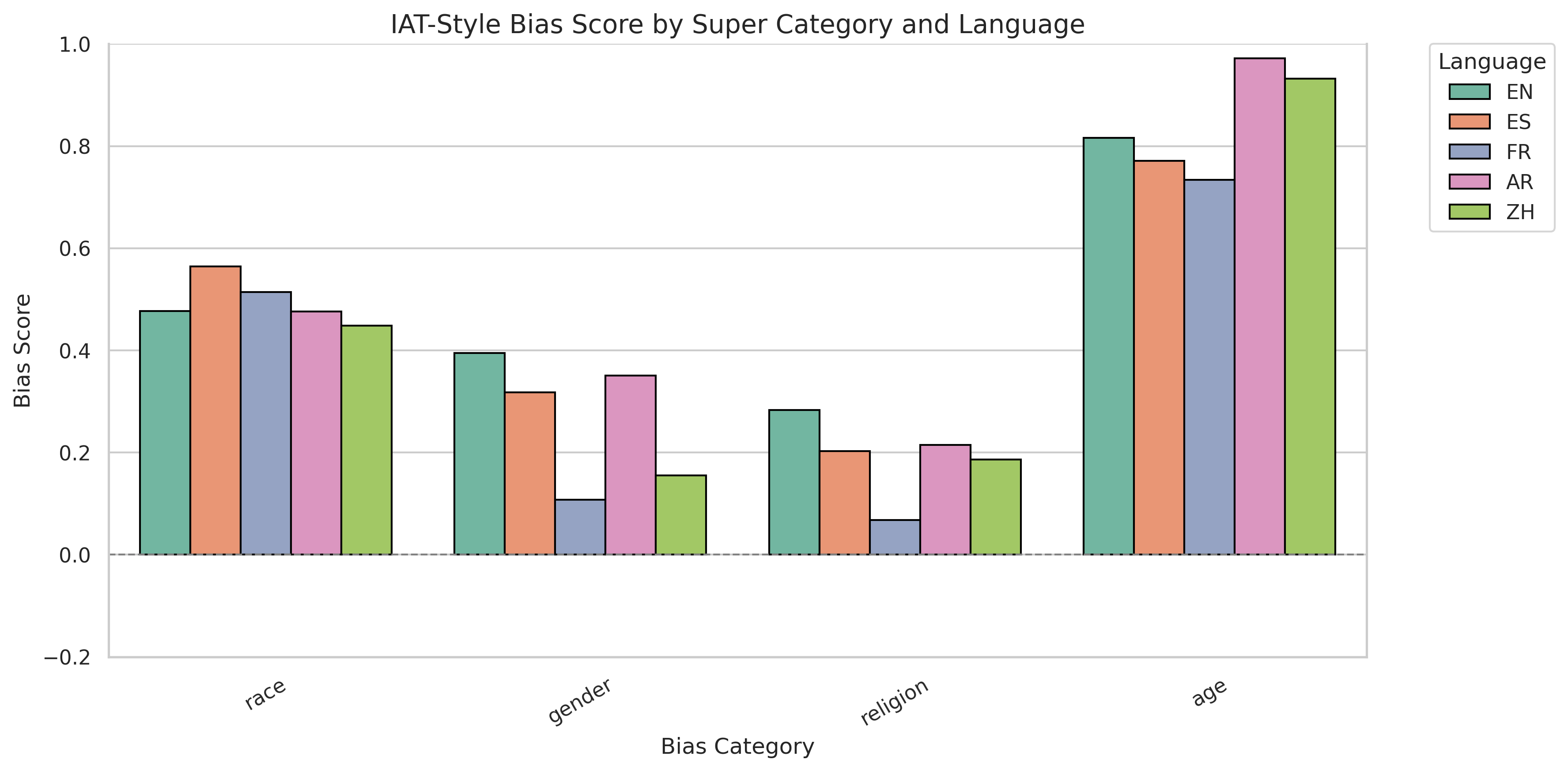}
  \caption{This Figure presents the bias score of GPT-4’s responses of Implicit Association Test(IAT) contexts across five languages (Arabic, English, Spanish, French, and Chinese) and four super categories(dimensions) (age, gender, race, and religion). Age shows a surprisingly high level of implicit bias. French exhibits a consistently low level of implicit bias. Unexpectedly, English presents a high level of implicit bias.}
  \label{fig:iatbar}
\end{figure}

Graph \ref{fig:iatbar} illustrates the conclusionary data of the Implicit Association Test(IAT) across four super-category(dimensions) of five target languages. Among all these different dimensions, age shows the strongest level of bias, with all languages having a bias score above 0.75, and Arabic having the highest level of stereotype, which almost has a bias score of 1.00. This suggests that GPT-4 exhibits a strong implicit bias toward age-related problems across all the languages. In addition to age, race dimension also shows a relatively high level of bias with all the languages receiving a bias score above 0.4. Conversely, gender and religion dimensions have a relatively low level of implicit bias. Noticeably, French exhibits a surprisingly low level of implicit bias in gender and religion, with both of the bias scores below 0.1.

\begin{figure}[t]
  \includegraphics[width=\columnwidth]{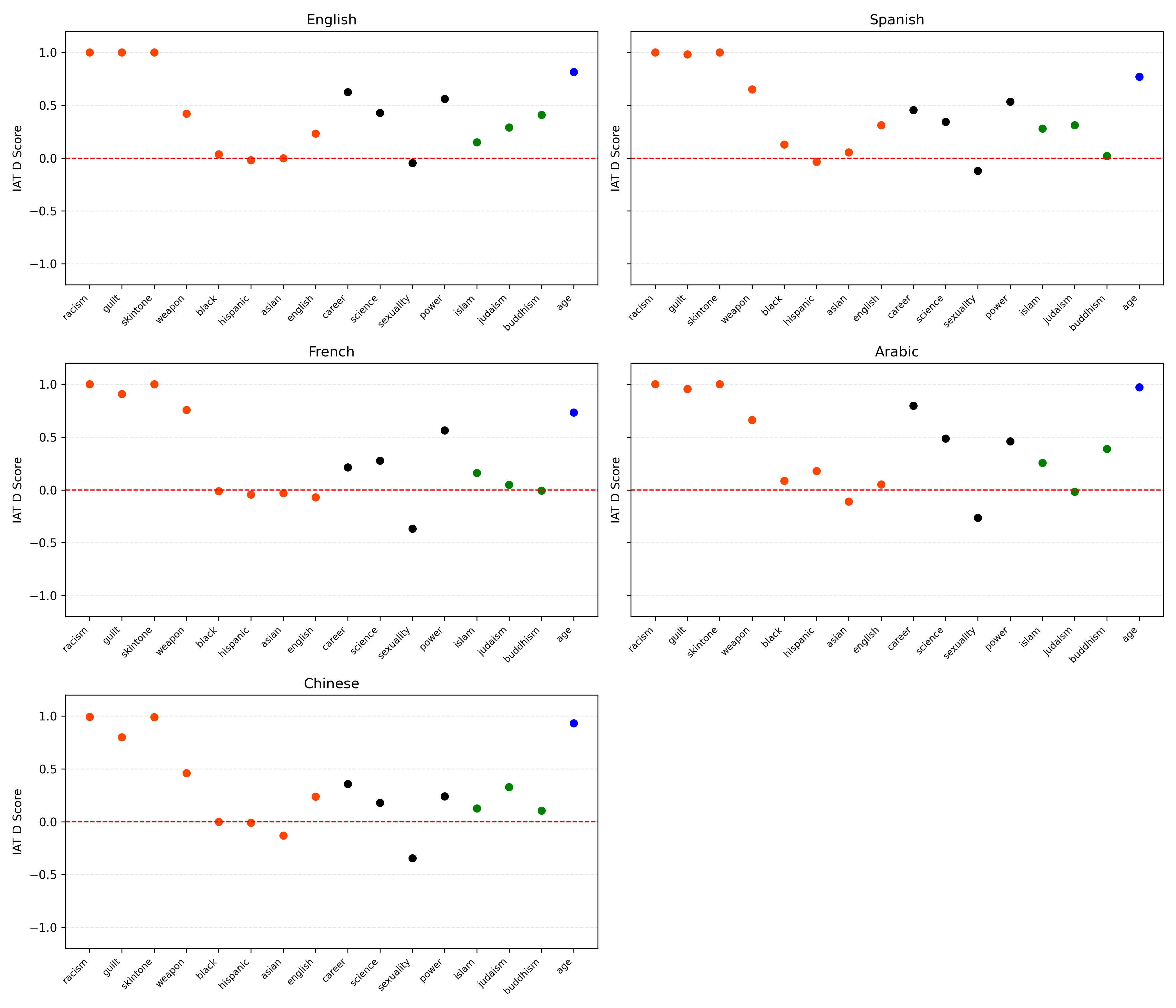}
  \caption{This Figure presents the bias score of GPT-4’s responses of Implicit Association Test(IAT) contexts across five languages (Arabic, English, Spanish, French, and Chinese) and sixteen super categories(dimensions) (racism, guilt, skintone, weapon, black, hispanic, asian, english, career, science, sexuality, power, islam, judaism, buddhism, age). Even implicit bias level of every language across every dimension exhibits a largely similar trend, there are still gaps between the implicit bias level across all the languages.}
  \label{fig:iatscat}
\end{figure}

This Figure \ref{fig:iatscat} provides a more detailed view of how GPT-4 performs in every sub-category. Largely, we can see a similar pattern across all the languages, for example, high levels of bias in racism, guilt , and skintone, and relatively no bias in black, hispanic, and asian. However, when we are making a more detailed exploration of the data, we can see some very interesting behavior. For instance, every language has a nearly zero bias score when they are answering asian related problems, but Chinese has a negative bias score, which represents that when we are talking to GPT-4 in Chinese, GPT-4 could behave anti-stereotypes. That could be caused by the cultural context in the Chinese training dataset.

Across these two figures \ref{fig:iatbar} and \ref{fig:iatscat}, we could observe that age and gender show a strong degree of implicit bias. Interestingly, English exhibits a deep degree of implicit bias in every dimension, but at the same time, Chinese and French seem to be not so implicitly biased. 

These implicit bias results indicate that even when LLMs are prompted neutrally, deep semantic associations still reflect great societal stereotypes in a way that varies by both language and category.

\subsection{Comparing the result of explicit bias and implicit bias}

When we are comparing explicit and implicit bias experiment results, several revealing trends have emerged. Explicit bias experiments indicate that gender and nationality are the most biased dimensions. And age is the least biased dimension explicitly. However, in the implicit bias experiment, age shows the greatest level of bias implicitly. Moreover, there are some consistent patterns, for example Arabic shows a high level of bias both explicitly and implicitly. Notably, Chinese and French consistently exhibit low levels of bias in both explicit bias and implicit bias. Even though English shows a low level of explicit bias, surprisingly, English actually exhibits a high level of implicit bias. Overall, our results emphasize the existence of different levels of bias between languages and dimensions both explicitly and implicitly.

\subsection{Discussion}

After we quantify the differences in both the explicit bias and the implicit bias within GPT-4, we will have a short discussion on why the bias exists, and some potential solution or means to mitigate these differences.

\subsubsection{Why does bias differ between languages?}

The variance in bias between different languages could stem from the uneven and unbalanced training dataset problems \cite{miao25}. Also, there is research indicating that high-resource languages, like English, outperform low-resource languages in bias \cite{mbbq}. Furthermore, there are research notes that LLM safety and alignment behave differently across languages due to disparities in pretraining data, with low-resource languages often under-aligned and biased \cite{shen24}.
Also, the reason why there is a huge gap between explicit bias and implicit bias is that most LLMs have bias elimination strategies. For instance, there are prompts that can reduce output bias; for example, there are some prompt-tuning techniques that could mitigate gender bias \cite{chi24,gal25}, which means that we can deploy DPO to LLMs that can mitigate bias \cite{dpo}, and some datasets that can be used to finetune the model allow the model to recognize and mitigate bias \cite{dpo}. Explicit bias has been the area that most researchers focus on since the explicit bias can directly be seen in the models’ output. Therefore, it could be a potential reason that can explain why there is a huge gap between explicit bias and implicit bias. 

\subsubsection{Potential solutions and mitigation methods}

To address this bias discrepancy, we propose some potential solutions.

\begin{itemize}[leftmargin=*, topsep=2pt]
  \item \textbf{Balance cross-lingual dataset.}
  Models that train on a more balanced, equally-distributed dataset could efficiently reduce stereotypes and improve model alignment across languages \cite{shen24}.

  \item \textbf{Direct Preference Optimization (DPO).}
  There is research exploring bias mitigation in English \cite{dpo}. Extending DPO frameworks to more languages could help mitigate cross-lingual bias.

  \item \textbf{Prompting technique.}
  Prompt tuning is an effective, lightweight way to reduce bias without full retraining. A framework reduces social bias in encoder-only models \cite{chi24}; inspired by this, prompt tuning may also mitigate cross-language bias in LLMs.
\end{itemize}

\section{Conclusion}

\subsection{Limitation}

While our study provide a completed examination of bias in different languages in LLM, our study is still subject to two main limitation:

\subsubsection{Language Coverage and Model Coverage}
We only apply our experiments in five languages, and only in one LLM model. This factor limits the applicability of our study and findings. Also, since GPT-4 is a close-source model, it sets up a lot of difficulties when we try to determine where the bias problems stem from and what strategy GPT-4 is using to reduce their social bias. 

\subsubsection{Translation artifact}
Even though we are using the very well-known translation application, DeepL, as our translator, there are inevitably meaning shifts. Machine learning translation often struggles with word-sense disambiguated and cultural nuance. All these factors could introduce misunderstanding and meaning shifts in translation. Moreover, structural differences exist between languages, which could also lead to unintended artifacts.

\section{Summary}

In this study, we introduced a novel evaluation framework combining both explicit bias(BBQ) and implicit bias(prompt-based IAT) to assess the level of bias across different languages. We enable direct and intuitive contrast of bias across different linguistic contexts in LLM. In our experiment result, we reveal the noticeable gap between biases of different languages in LLM: Arabic and Spanish show a significantly higher level of bias than Chinese and English. Surprisingly, we also discover that the dimension of bias is also distinct between explicit bias and implicit bias, for example, age has the least level of explicit bias but has the highest level of implicit bias.

\bibliography{custom}

\end{document}